\pdfoutput=1
\documentclass[conference]{IEEEtran}
\IEEEoverridecommandlockouts
\usepackage{cite}
\usepackage[numbers]{natbib}
\usepackage{amsmath,amssymb,amsfonts}
\usepackage{algorithmic}
\usepackage{graphicx}
\usepackage{textcomp}
\usepackage[inline]{enumitem}
\usepackage[table]{xcolor}
\usepackage{todonotes}
\usepackage{hyperref}
\usepackage{booktabs}
\usepackage{diagbox}
\usepackage[frozencache=true,cachedir=minted-cache]{minted} 
\usepackage[T1]{fontenc} 
\usepackage{sourcecodepro} 

\definecolor{myred}{rgb}{1, 0.941, 0.941}
\definecolor{mygreen}{rgb}{0.941, 0.9725, 0.949}
\definecolor{myyellow}{rgb}{1, 0.988, 0.941}

\setminted{fontsize=\small}
\setminted{autogobble=true}

\def\codeinline#1{\mintinline{text}{#1}}
\def\pyinline#1{\mintinline{python}{#1}}

\newcommand{\highlight}[1]{\begin{framed}%
  \noindent\emph{#1}
\end{framed}}

\def\mybar#1{
  {\rule{#1in}{1ex}}}

\def\mybar#1{
  {\rule{#1in}{1ex}}}

\def\BibTeX{{\rm B\kern-.05em{\sc i\kern-.025em b}\kern-.08em
    T\kern-.1667em\lower.7ex\hbox{E}\kern-.125emX}}

\begin{document}
\title{The Prevalence of Code Smells in Machine Learning projects}

\author{\IEEEauthorblockN{Bart van Oort$^{1,2}$, Lu\'{i}s Cruz$^2$, Maur\'{i}cio Aniche$^2$, Arie van Deursen$^2$}
\IEEEauthorblockA{\textit{Delft University of Technology} \\
\textit{$^1$~AI for Fintech Research, ING} \\
$^2$~Delft, Netherlands \\
bart.van.oort@ing.com, \{l.cruz, m.f.aniche, arie.vandeursen\}@tudelft.nl}
}

\maketitle

\thispagestyle{plain}
\pagestyle{plain}

\begin{abstract}
Artificial Intelligence (AI) and Machine Learning (ML) are pervasive in the current computer science landscape. Yet, there still exists a lack of software engineering experience and best practices in this field. One such best practice, static code analysis, can be used to find code smells, i.e., (potential) defects in the source code, refactoring opportunities, and violations of common coding standards. Our research set out to discover the most prevalent code smells in ML projects. We gathered a dataset of 74 open-source ML projects, installed their dependencies and ran Pylint on them. This resulted in a top 20 of all detected code smells, per category. Manual analysis of these smells mainly showed that code duplication is widespread and that the PEP8 convention for identifier naming style may not always be applicable to ML code due to its resemblance with mathematical notation. More interestingly, however, we found several major obstructions to the maintainability and reproducibility of ML projects, primarily related to the dependency management of Python projects. We also found that Pylint cannot reliably check for correct usage of imported dependencies, including prominent ML libraries such as PyTorch.
\end{abstract}


\section{Introduction} \label{sec:introduction}

Artificial Intelligence (AI) and Machine Learning (ML) are pervasive in the current landscape of computer science. Companies such as Facebook, Google, Nvidia and ING are making use of AI and ML for a plethora of tasks that are difficult (if not impossible) to describe using traditional Software Engineering (SE)~\cite{haakman2020ai,sculley2015hidden,amershi2019software,gonzalez2020mluniverse}. Examples include facial recognition \& recomposition, natural language processing, real-time video transformation, detection of medical anomalies and intercepting fraudulent financial transactions.

Yet, as \citet{sculley2015hidden} wrote in their 2015 paper on the hidden technical debt in ML systems at Google, \textit{``only a small fraction of real-world ML systems is composed of the ML code. (...) The required surrounding infrastructure is vast and complex.''} This is also in part what leads \citet{menzies2020lawsofSEforAI} to predict that the future of software will be a rich and powerful mix of ideas from both SE and AI. \citeauthor{menzies2020lawsofSEforAI} also advocates for more SE experience in the field of AI and ML, stating that poor SE leads to poor AI while better SE leads to better AI \cite{menzies2020lawsofSEforAI}. The data scientists that write AI / ML code often come from non-SE backgrounds where SE best practices are unknown \cite{simmons2020dsstandards}.

One such SE best practice is the practice of static code analysis to find (potential) defects in the source code, refactoring opportunities and violations of common coding standards, which we amalgamate into `code smells' for the rest of this paper. Research has shown that the attributes of quality most affected by code smells
are maintainability, understandability and complexity, and that early detection of code smells reduces the cost of maintenance \cite{lacerda2020smellsSLR}.

With a focus on the maintainability and reproducibility of ML projects, the goal of our research is therefore to apply static code analysis to applications of ML, in an attempt to uncover the frequency of code smells in these projects and list the most prevalent code smells. Thus, we formulate the following research question: \textit{What are the most prevalent code smells in Machine Learning code?}

    

The main contributions of this paper are:
\begin{itemize}
    \item An empirical study on the prevalence of code smells in 74 Python ML projects.
    \item A dataset of 74 ML projects and an open-source tool to perform simultaneous static code analysis on all of these projects.
\end{itemize}

\section{Related Work}

Several studies have investigated linting and static code analysis of non-ML projects \cite{tomasdottir2020eslint,omari2019usesPylint,bafatakis2019usesPylint,chen2018pysmells}. \citet{tomasdottir2020eslint} researched why JavaScript (JS) developers use linters and how they tend to configure them. They found that maintaining code consistency, preventing errors, saving discussion time and avoiding complex code were among the top reasons why JS developers use linters. They also found that JS developers commonly stick with already existing preset linting configurations. \citet{vassallo2020asats} found a similar result; among other results, they found that developers are often unwilling to configure automatic static analysis tools (ASATs) and emphasise \textit{``the necessity to improve existing strategies for the selection and prioritisation of ASATs warnings that are shown to developers.''}

Within the Python ecosystem, \citet{chen2018pysmells} investigated the detection and prevalence of code smells in 106 Python projects with the most stars on GitHub. They found that long parameter lists and long methods were more prevalent than other code smells. \citet{omari2019usesPylint} used Pylint to analyse the code quality of a dataset of large Python projects. Furthermore, \citet{bafatakis2019usesPylint} used Pylint to investigate the Python coding style compliance of StackOverflow answers.

Within the Machine Learning ecosystem, we only found one paper by \citet{simmons2020dsstandards} that performed static code analysis on a large dataset of Data Science (DS) projects. They also analysed non-DS projects with the goal of comparing the code quality and coding standard conformance of (open-source) DS projects versus non-DS projects, using Pylint in its default configuration as a metric. They sourced their DS projects from \citet{biswas2019dsdataset}, who in 2019 published a dataset of 1558 \textit{``mature Github projects that develop Python software for Data Science tasks.''}. Aside from applications of ML, it also includes ML libraries and tools.

Our study differs from \cite{simmons2020dsstandards} in that we do not compare against non-DS projects and in that we do not solely focus on the adherence to coding standards as \cite{simmons2020dsstandards} does. Our primary focus lies more on investigating obstructions to the maintainability and reproducibility of ML projects, which includes coding standards violations, but also entails recognising refactoring opportunities and other code smells \cite{lacerda2020smellsSLR}. Moreover, we solely focus on applications of ML, and leave ML libraries and tools out of scope. We argue that the underlying nature of ML libraries and tools is very different from ML applications, and thus different results are expected when studied separately.

Furthermore, \citet{simmons2020dsstandards} simplified the installation of the projects' dependencies by using \codeinline{findimports}\footnote{\url{https://pypi.org/project/findimports/}} to resolve all imports used in the projects, instead of relying on what projects' authors defined in their repositories, noting that \textit{``it was impractical to reliably determine and install dependencies for the projects analysed.''} However, if there is an inherent difficulty in resolving these dependencies within Python projects, then that is in itself an obstruction to the reproducibility and maintainability of these projects. Hence, we investigate this in our study.

\section{Methodology}

For this paper, we performed an empirical study on the prevalence of code smells in ML code. We collected a dataset of 74 ML projects and implemented a tool to set these projects up with their dependencies in order to replicate their execution environment. It then runs Pylint with its default configuration on all projects in the dataset, collecting and counting the detected code smells. The tool and dataset are both open-source and can be found on GitLab\footnote{\url{https://gitlab.com/bvobart/python-ml-analysis}}.

\begin{figure*}[htbp]
    \centering
    \includegraphics[width=0.9\textwidth]{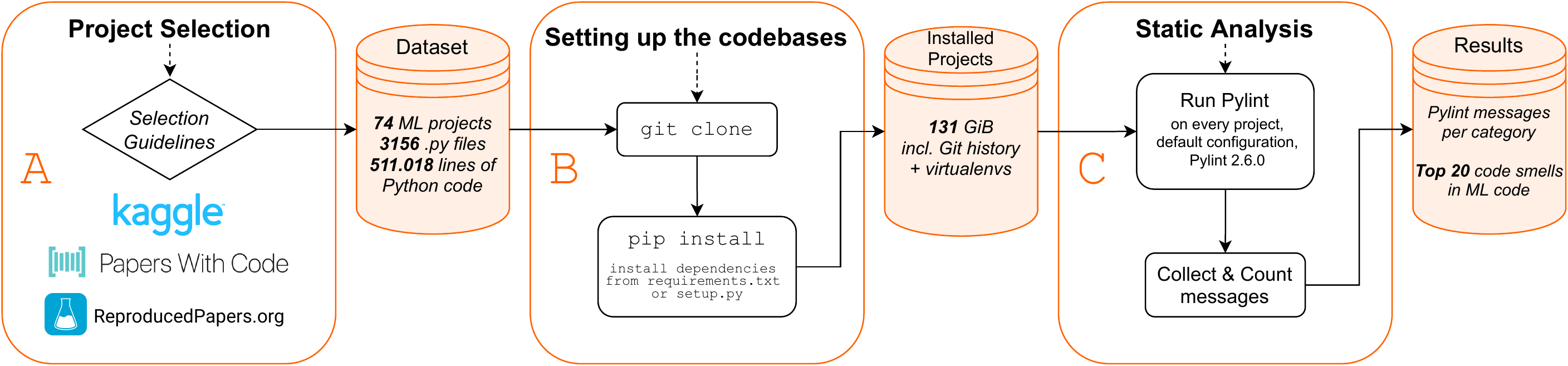}
    \caption{Methodology Diagram.}
    \label{fig:methodology}
    \vspace{1mm}
\end{figure*}

Our empirical study follows the methodology illustrated in Figure~\ref{fig:methodology}. It comprises three main steps, namely: A) project selection, B) setting up the codebases, and C) static analysis.

\subsection{Project Selection}

In total, our collected dataset comprises 74 ML projects; 32 projects come from finished Kaggle competitions, 38 from \url{paperswithcode.com} (of which 25 projects were from the Google-affiliated DeepMind), and 4 from \url{reproducedpapers.org}. It includes projects from academic papers, (student) reproductions, prize money awarding Kaggle competitions, as well as industry players such as Facebook, Nvidia and DeepMind. The dataset defines a list of Git repository URLs and allows for customising the dependencies of particular projects, when they have not been properly defined in their respective repository. We elaborate on a number of characteristics of our dataset in Section \ref{sec:methodology:selection:characteristics}, but first, we explain how we collected the projects in the dataset and what guidelines were used for doing so.

We aim for this dataset to be a systematically gathered set of projects, representative of the current, real-world state of ML and AI projects. To this end, we have created a set of guidelines for the inclusion of projects in the dataset, which can be found below. Each project included in the dataset\dots

\begin{enumerate}
    \item \dots must be hosted in an open-source Git repository.
    \item \dots must be written in Python 3.
    \item \dots must contain pure Python code and does not consist purely of Jupyter Notebooks. More specifically, a project should contain \textit{either} a) at least 200 lines of pure Python code, even if the rest of the code is embedded in Jupyter Notebooks, \textit{or} b) more lines of pure Python code than there are lines of Python code in all Jupyter notebooks of that project.
    \item \dots must implement an ML or AI model and may not be a library or tool for use in ML projects.
    \item \dots must be considered `deliverable', i.e., \textit{either} a) the project is part of or accompanies a published academic paper, \textit{or} b) the project has been submitted to \url{paperswithcode.com}, \url{reproducedpapers.org} or a Kaggle competition (which has finished and declared the winners at the time of considering the project).
\end{enumerate}

The first guideline limits our scope to open-source projects, as these are openly available to download and analyse.

The second and third guideline stem from a technical limitations, as Pylint only supports Python 3 and is only able to analyse pure Python files. Jupyter Notebooks are essentially JSON files, containing `cells' with code in Markdown, Python, Julia, or a small selection of other different languages. While it is technically possible to convert the Python code embedded in these notebooks to pure Python files using a tool such as \codeinline{nbconvert}, the produced code has a slightly different style than general Python modules, which invalidates certain Pylint rules. For example, the Pylint messages \codeinline{pointlesss-statement}, \codeinline{expression-not-assigned} and \codeinline{wrong-import-statement} produce false positives in notebook-style code. Due to the lack of direct Pylint support for Jupyter Notebooks and since we do not want to selectively disable Pylint rules for notebook-extracted code as opposed to pure Python code, we decided to exclude projects that purely contain Jupyter Notebooks from our dataset. The minimum of 200 lines of pure Python code in the presence of larger Jupyter Notebooks was chosen such that this code is likely not to be purely utility code, but also contain part of the ML code.


The fourth guideline embodies that we are interested in analysing applications of ML rather than libraries used in their development, such as \codeinline{tensorflow}, \codeinline{pandas}, or \codeinline{sklearn}.

The fifth and final guideline focuses on avoiding toy projects, unfinished projects, or projects still under development. 

\subsubsection{Dataset characteristics} \label{sec:methodology:selection:characteristics}

We measured several general characteristics of every project, which can be found in \autoref{tab:methodology:characteristics}. The 74 projects in the dataset contain a total of 3156 pure Python files, amounting to 511.018 lines of Python code, including empty lines. The median project has 17 pure Python files with 2848.5 lines of code, resulting in an average of 157.3 lines of Python code per file. The smallest project contained a single file with 58 lines of Python, while the largest project had 78572 lines of Python code across 229 files. This project was found to be embedding the code of several dependencies in its repository, multiple times.

\begin{table*}
\caption{Characteristics of our dataset of 74 ML projects}
\label{tab:methodology:characteristics}
\centering
\begin{tabular}{lrrrrrrr}
\toprule
\textit{Characteristic}                    & \textit{Min} & \textit{Q1}     & \textit{Median}     & \textit{Q3}     & \textit{Max}   & \textit{Mean}   & \textit{Std Dev.} \\ 
\midrule

Number of pure Python files                  & 1   & 8      & 17     & 36     & 730   & 43   & 95    \\
Number of Jupyter Notebook files             & 0   & 0      & 0      & 1      & 52    & 3    & 8     \\
Lines of Python code                         & 58  & 1449   & 2849  & 5243   & 78572 & 6906 & 13568 \\
Lines of Jupyter Notebook Python             & 0   & 0      & 0      & 197    & 43387 & 1008 & 5115  \\
Avg. lines of Python code per Python file    & 58  & 108    & 157  & 214    & 1151  & 193  & 155 \\
Avg. lines of Jupyter Notebook Python per Jupyter Notebook file & 12 & 73 & 129 & 223 & 1610 & 256 & 335 \\
\bottomrule
\end{tabular}
\vspace{1mm}
\end{table*}



\subsection{Setting up the codebases} \label{sec:methodology:setting-up}

In this step, performed by our analysis tool, for each project, we clone the latest version of the project's Git repository, create a virtual environment in it, ensure that there exists a file that specifies any necessary dependencies and then install those dependencies into the virtual environment. These need to be installed, such that our static analysis tool of choice is able to check whether imports resolve correctly and whether the imported libraries are used correctly. This is particularly interesting in the case of ML, as \citet{sculley2015hidden} noted that ML projects have a high degree of glue code and so make extensive use of libraries. In total, the folder containing all of the 74 cloned projects from our dataset along with the accompanying virtual environments with their installed dependencies, amounts to 131 GiB.

Python projects can specify and install their dependencies in a variety of ways. The most common way to install a Python dependency is to use \codeinline{pip}, the package manager that is installed alongside Python. It is common convention to specify a Python project's list of dependencies in a requirements file called \codeinline{requirements.txt}, which is conventionally placed at the root of the project's source code repository. It is also possible to specify a \codeinline{setup.py} file which allows the project to be built into a Python package, ready to be published to PyPI, \codeinline{pip}'s default package index. \codeinline{pip} can be configured to use other package indexes, but by default it can only install packages from PyPI, or directly from source through a Git repository URL or a local folder with a \codeinline{setup.py}.

However, there are also other package managers / dependency management solutions such as Conda, Poetry, \codeinline{pip-tools} and Pipenv, with the latter being directly endorsed in Python's Packaging User Guide \cite{python-packaging-guide}. These tools each have their own way of specifying dependencies and -- especially in Conda's case -- may use additional package indexes to PyPI, which makes resolving these dependencies difficult. It is possible to use \codeinline{pip freeze > requirements.txt}, which collects all Python packages and their exact versions installed in the current Python environment (disregarding by which means these packages were installed) and outputs them to a \codeinline{requirements.txt} file. This approach is flawed though, as we explain in Section \ref{sec:discussion:dependencies}.


Our analysis tool currently only supports installing dependencies with \codeinline{pip} and expects a \codeinline{requirements.txt} or \codeinline{setup.py} file in their conventional location. The dataset also supports specifying a custom path to a \codeinline{requirements.txt} file, or alternatively, the contents of a custom \codeinline{requirements.txt} file for a project in the dataset. It is also possible to specify extra requirements that need to be installed \textit{after} installing the dependencies from the requirements file. This is necessary for, e.g., Nvidia's Apex library, which depends on PyTorch; when trying to run \codeinline{pip install} on a requirements file containing both PyTorch and Apex's Git repository URL (no matter the order), the installation of Apex fails because PyTorch is not yet installed. Only for projects that do not have a \codeinline{requirements.txt} file, nor a manually defined one, our analysis tool uses \codeinline{pipreqs}\footnote{\url{https://github.com/bndr/pipreqs}} to generate a \codeinline{requirements.txt} file based on the libraries imported in the code.

Our analysis tool currently does not support using Conda, Poetry or Pipenv for resolving and installing dependencies. We therefore had to exclude one project that used Poetry and two projects that used Conda and solely specified a Conda \codeinline{environment.yml} file, but no \codeinline{requirements.txt} or \codeinline{setup.py}. No projects that we came across were using Pipenv.

\subsection{Static Analysis}

This step is also performed by our analysis tool and concerns running the static code analysis tool Pylint (version 2.6.0) in its default configuration on all pure Python files in each project (but not on any of the dependencies). We choose Pylint for static code analysis as it is widely used and widely accepted in the Python community, as well as being highly configurable \cite{simmons2020dsstandards,bafatakis2019usesPylint}. It is also well integrated into IDEs such as PyCharm and VS Code.
Furthermore, \citet{bafatakis2019usesPylint} used it to measure coding style compliance in StackOverflow answers, \citet{omari2019usesPylint} used it as a metric for the code quality of open-source Python projects, and \citet{simmons2020dsstandards} used it in their code quality comparison between DS and non-DS Python projects.

Pylint provides an extensive set of messages, not only for stylistic issues, but also for issues regarding programming conventions, possible refactorings and other logical code smells. While Pylint is very configurable, we chose to use Pylint's default configuration as it reflects the community standards, similar to \citet{simmons2020dsstandards}.


The code smells that Pylint reports are each identified by a symbol, such as \codeinline{bad-indentation} or \codeinline{import-error}, which is also how we refer to specific Pylint messages in this paper. Furthermore, these messages are divided into five categories (message types), which we describe below. The italic text is how Pylint describes the category.

\begin{itemize}
    \item \textbf{Convention} -- \textit{for programming standard violation} -- Messages in this category show violations of primarily code style conventions, as well as documentation conventions and Pythonic programming conventions.
    \item \textbf{Refactor} -- \textit{for bad code smell} -- Messages in this category indicate that the smelly code should be refactored. 
    \item \textbf{Warning} -- \textit{for Python specific problems} -- This category includes many generic and Python-specific linting messages.
    \item \textbf{Error} -- \textit{for probable bugs in the code} -- Messages in this category indicate problems in the code that are very likely to cause run-time problems.
    \item \textbf{Fatal} -- \textit{if an error occurred which prevented Pylint from doing
    further processing}.
\end{itemize}


Cloning and installing all projects, even though this is performed automatically by the tool, is the most time-consuming part of the analysis -- it takes roughly three hours. With all projects already cloned and their dependencies already installed, the analysis of all 74 projects took 8m 53s using 12 threads on an Intel\textsuperscript{\textregistered{}} Core\texttrademark~i7-8750H processor.

Eight projects contained code that caused Pylint to crash during analysis, so we excluded these from the 82 projects we originally had in the dataset, bringing the total to 74. Several issues have been filed about this, including one by this paper's first author\footnote{\url{https://github.com/PyCQA/pylint/issues/3986}}. This bug has since been fixed.

\section{Results} \label{sec:results}

Applying our methodology, we collected, installed, and analysed 74 ML projects. In this section, we present our results and answer to the research question posed in the introduction:

\begin{itemize}
    \item \textbf{RQ} -- \textit{What are the most prevalent code smells in Machine Learning code?}
\end{itemize}

To answer this, we first analysed the distribution of the amount of code smells per Pylint category per project, of which the characteristics can be found in \autoref{tab:results:msgs-per-category}. The table shows the minimum, maximum, mean and median number of messages reported by Pylint for each category, as well as the 25th percentile (Q1), 75th percentile (Q3), and standard deviation (Std. Dev.). We use the median as the main measure of central tendency.

Our results show that Pylint messages in the Warning category are the most prevalent -- the median project has 356 warnings -- closely followed by messages in the Convention category with 226 messages for the median project. Messages in the Refactor and Error categories are less prevalent; respectively 49 and 56 such messages for the median project. However, especially given the Error category is meant for messages that show ``probable bugs'', this is an interesting observation. Even more interesting, there was \textit{no} project for which Pylint reported \textit{no} error messages.

\begin{table}
\centering
\caption{Distribution of Pylint messages per category per project.}
\label{tab:results:msgs-per-category}
    \resizebox{\linewidth}{!}{\begin{tabular}{lrrrrrrr}
    \toprule
    \textit{Category}       & \textit{Min}  & \textit{Q1}  & \textit{Median}  & \textit{Q3}   & \textit{Max}            & \textit{Mean}  & \textit{Std Dev.} \\ \midrule
    Convention              & 2   & 57  & 226 & 799 & 9501  & 708   & 1361  \\
    Refactor                & 0   & 26  & 49  & 140 & 2437  & 183   & 433   \\
    Warning                 & 11  & 74  & 356 & 824 & 14263 & 814   & 1826  \\
    Error                   & 1   & 18  & 56  & 125 & 1696  & 129   & 234   \\
    Fatal                   & 0   & 0   & 0   & 0   & 0     & 0     & 0       \\
    \bottomrule
    \end{tabular}}
\end{table}

As a more direct answer to this research question, we measured across all projects in our dataset what the top 20 code smells per category are that Pylint reported, see \autoref{tab:results:top20-per-category}. The top 10 messages that Pylint reported, disregarding category, are in \autoref{tab:results:top10-all}.

\begin{table}
\centering
\vspace{-2em}
\caption{Top 10 code smells overall as detected by Pylint.}
\label{tab:results:top10-all}
    \begin{tabular}{rlrl}
    \toprule
    \textit{\#} & \textit{Smell} & \textit{Frequency} &  \\
    \midrule
    \textit{1} & unused-wildcard-import & 26307 & \mybar{0.263}\\
    \textit{2} & bad-indentation & 19921        & \mybar{0.199}\\
    \textit{3} & invalid-name & 19905           & \mybar{0.199}\\
    \textit{4} & line-too-long & 10321          & \mybar{0.103}\\
    \textit{5} & missing-function-docstring & 6444  & \mybar{0.064} \\
    \textit{6} & no-member & 5860                   & \mybar{0.059}\\
    \textit{7} & duplicate-code & 4649              & \mybar{0.045}\\
    \textit{8} & trailing-whitespace & 4477         & \mybar{0.045}\\
    \textit{9} & redefined-outer-name & 2548        & \mybar{0.025}\\
    \textit{10} & missing-module-docstring & 2504   & \mybar{0.025}\\
    \bottomrule
    \end{tabular}
\end{table}

\begin{table*}[htbp]
\centering
\caption{Top 20 messages per category reported by Pylint along with how often they were counted across all projects.}
\label{tab:results:top20-per-category}
\resizebox{\textwidth}{!}{%
\begin{tabular}{llr|lr|lr|lr}
\toprule
            & \multicolumn{2}{l|}{\textbf{Convention}}          & \multicolumn{2}{l|}{\textbf{Refactor}}              & \multicolumn{2}{l|}{\textbf{Warning}}                & \multicolumn{2}{l}{\textbf{Error}}               \\
            & \textit{Symbol}                 & \textit{Count} & \textit{Symbol}                   & \textit{Count} & \textit{Symbol}                    & \textit{Count} & \textit{Symbol}                 & \textit{Count} \\ \midrule
\textit{1}  & invalid-name                    & 19905          & duplicate-code                    & 4649           & unused-wildcard-import             & 26307          & no-member                       & 5860           \\
\textit{2}  & line-too-long                   & 10321          & too-many-arguments                & 2158           & bad-indentation                    & 19921          & import-error                    & 1750           \\
\textit{3}  & missing-function-docstring & 6444                & super-with-arguments         & 1802                & redefined-outer-name          & 2548                & undefined-variable              & 471            \\
\textit{4}  & trailing-whitespace        & 4477                & too-many-locals              & 1456                & unused-import                 & 2321                & not-callable                    & 397            \\
\textit{5}  & missing-module-docstring   & 2504                & too-many-instance-attributes  & 658                & arguments-differ              & 1678                & no-name-in-module               & 326            \\
\textit{6}  & wrong-import-position      & 2286                & no-else-return                & 509                & unused-variable                & 986                & no-value-for-parameter          & 168            \\
\textit{7}  & missing-class-docstring    & 2060                & too-few-public-methods        & 438                & attribute-defined-outside-init & 962                & function-redefined              & 100            \\
\textit{8}  & wrong-import-order         & 1750                & no-self-use                   & 422                & unused-argument                & 902                & unsubscriptable-object          & 100            \\
\textit{9}  & ungrouped-imports           & 367                & useless-object-inheritance    & 351                & abstract-method                & 841                & bad-option-value                & 94             \\
\textit{10} & import-outside-toplevel     & 285                & too-many-statements           & 265                & redefined-builtin              & 536                & unexpected-keyword-arg          & 84             \\
\textit{11} & consider-using-enumerate    & 256                & too-many-branches             & 218                & dangerous-default-value        & 447                & relative-beyond-top-level       & 68             \\
\textit{12} & missing-docstring           & 246                & cyclic-import                 & 108                & reimported                     & 322                & assignment-from-no-return       & 27             \\
\textit{13} & superfluous-parens          & 244                & inconsistent-return-statements & 71                & wildcard-import                & 297                & bad-super-call                  & 21             \\
\textit{14} & missing-final-newline       & 236                & unnecessary-comprehension      & 48                & logging-format-interpolation   & 288                & redundant-keyword-arg           & 19             \\
\textit{15} & multiple-statements         & 218                & chained-comparison             & 46                & pointless-statement            & 253                & too-many-function-args          & 18             \\
\textit{16} & trailing-newlines           & 176                & consider-using-in              & 45                & fixme                          & 247                & invalid-unary-operand-type      & 17             \\
\textit{17} & bad-whitespace              & 166                & simplifiable-if-expression     & 34                & protected-access               & 223                & no-self-argument                & 9              \\
\textit{18} & unidiomatic-typecheck        & 78                & literal-comparison             & 30                & logging-fstring-interpolation  & 110                & misplaced-bare-raise            & 9              \\
\textit{19} & singleton-comparison         & 69                & too-many-nested-blocks         & 26                & f-string-without-interpolation & 105                & no-method-argument              & 6              \\
\textit{20} & multiple-imports             & 53                & no-else-raise                  & 24                & pointless-string-statement     & 101                & access-member-before-definition & 6             \\
\bottomrule
\end{tabular}%
}

\end{table*}

\textbf{Convention} --
In this category we found that invalid naming, missing documentation (\codeinline{missing-function-docstring}, \codeinline{missing-module-docstring}, \codeinline{missing-class-docstring} and \codeinline{missing-docstring}) and improper organisation of imports (\codeinline{wrong-import-position}, \codeinline{wrong-import-order}, \codeinline{ungrouped-imports}, \codeinline{import-outside-toplevel} and \codeinline{multiple-imports}) were the most commonly recognised code smells in the Convention category.

\textbf{Refactor} --
The most commonly recognised opportunities for refactoring pertained to duplicate code (4649), using too many arguments when defining a function or method (2158, \codeinline{too-many-arguments}), and using an old style for calling \pyinline{super} in the constructor of an inheriting class (1802, \codeinline{super-with-arguments}), instead of using the Python 3 style where no arguments to \pyinline{super} are necessary. It also shows that functions and classes are often too complex; Pylint reports 1456 functions that use too many local variables (\codeinline{too-many-locals}), 265 that are too long (\codeinline{too-many-statements}) and 218 that have too many branches, as well as 658 classes that have too many attributes on them (\codeinline{too-many-instance-attributes}).

\textbf{Warning} --
The most reported Warning messages, by far, are \codeinline{unused-wildcard-import} (26307) and \codeinline{bad-indentation} (19921). Code smells relating to import management, as already indicated in the Convention category, are also reflected in the Warning category with 26307 counts of unused wildcard imports, 2321 counts of unused imports, 322 counts of libraries that were imported multiple times in the same file (\codeinline{reimported}) and 297 counts of wildcard imports. Aside from unused imports, unused variables (986) and unused arguments (902) are also common. Having variables that redefine (shadow) function or variable names from an outer scope (\codeinline{redefined-outer-scope}) is also common with 2548 recognised cases, as is redefining Python's built-in global names (536, \codeinline{redefined-builtin}).

\textbf{Error} --
Finally, in the Error category, with 5860 counts, the \codeinline{no-member} message is the most prevalent, warning about the usage of non-existent attributes and methods on class instances and non-existent functions in Python modules. Import errors are the second most common with 1750 counts (i.e. on average 23.6 import errors per project), which are reported when a module (whether an external library or a module from a local file) contains imports that Pylint cannot resolve. The 326 \codeinline{no-name-in-module} messages are also related to these import problems, as they are emitted upon using a \pyinline{from X import Y} style import, where \codeinline{X} is resolved (so no import error is emitted), but \codeinline{Y} is not found. Furthermore, the use of undefined variables and attempting to call uncallable objects are also prevalent.

\section{Implications} \label{sec:discussion}

In this section, we discuss the implications of our results for ML developers. We start by elaborating upon the code smells that we have found to be most prevalent and continue with a discussion of problems regarding dependency management that we encountered while performing this research. We also argue how these problems affect the maintainability and reproducibility of the analysed ML projects.



\subsection{Explaining the most prevalent code smells} \label{sec:discussion:code-smells}
This section aims at providing an explanation for the prevalence of the most common code smells in our dataset of ML projects by investigating their occurrences.

\textbf{Error} --
Most interestingly, we found that there were \textit{zero} projects that had \textit{zero} messages in the Error category. Only the \codeinline{geomancer} project in the DeepMind research repository\footnote{\url{https://github.com/deepmind/deepmind-research}} had one error, namely a true positive \codeinline{no-name-in-module} error message in the project's test file.

We also found that \codeinline{no-member} and \codeinline{import-error} are the most reported code smells in this category. Upon manual inspection of these messages in several projects, we noticed that import errors have two primary causes, namely:
\begin{itemize}[leftmargin=*]
    \item \textbf{Bad specification of requirements} -- Using the \codeinline{kaggle-kuzushiji-recognition-2019} project as an example, we noticed that it was missing at least four dependencies in its (otherwise well-defined) requirements file. Imports of these missing dependencies were primarily found in the code of a dependency that the project's authors had copied into their repository for some slight customisations, but were also found in other scripts in the repository.
    \item \textbf{Pylint producing false positives on local imports} -- Taking the \codeinline{navigan}\footnote{\url{https://github.com/yandex-research/navigan}} project as an example, even though we manually fixed the import errors relating to badly specified requirements, there were 28 errors remaining. These errors come from unresolved imports from local modules, i.e., Python files in the repository. In Python, a local module \codeinline{utils.py} can be imported from other modules in the same directory using \pyinline{import utils} and it is recommended (but not necessary) to add an \codeinline{__init__.py} file to that directory to indicate that it is a Python package \cite{python-docs-initpy}. However, as a GitHub issue reports\footnote{\url{https://github.com/PyCQA/pylint/issues/3984}}, Pylint produces false positive import errors on local imports, but strangely \textit{not} when the \codeinline{__init__.py} file is \textit{not} present.
\end{itemize}

As for \codeinline{no-member} errors, in the \codeinline{kaggle-kuzushiji-recognition-2019} project -- which has 327 of them -- these were primarily caused by false positives from Pylint on the majority of -- if not all -- usages of the \codeinline{torch} library (i.e. PyTorch), including those of basic PyTorch functions like \pyinline{torch.as_tensor}, \pyinline{torch.tensor} and \pyinline{torch.max}. The project with the most \codeinline{no-member} errors, \codeinline{RSNA-STR-Pulmonary-Embolism-Detection} (1549), showed the same trend, as did \codeinline{DL-unet} and several other projects that we investigated. This is a known issue that has been reported to Pylint's Github repository\footnote{https://github.com/PyCQA/pylint/issues/\{\href{https://github.com/PyCQA/pylint/issues/3510}{3510}, \href{https://github.com/PyCQA/pylint/issues/2708}{2708}, \href{https://github.com/PyCQA/pylint/issues/2067}{2067}\}.} of which the essence goes back as far as 2013 with a similar problem in the use of NumPy\footnote{\url{https://github.com/PyCQA/pylint/issues/58}.}. The reason that is stated in these issues, is that Pylint has trouble extracting the members and type information of libraries that are backed by bindings with the C programming language.

\highlight{\textbf{Pylint cannot reliably check} for correct uses of \textbf{import statements}; both local imports, as well as imports from C-backed libraries such as PyTorch, suffer from a \textbf{high rate of false positives}.}

This is especially concerning in the context of ML, as the majority of ML libraries are backed by C (to make them performant). The fix that the Pylint developers propose in the relevant GitHub issues all entail (partially) disabling the \codeinline{no-member} rule, implying that Pylint cannot reliably check for correct uses of C-backed libraries. This is additionally concerning in the context of ML, as it has a high degree of glue code, i.e. code that is written to coerce data in and out of general-purpose libraries \cite{sculley2015hidden}.

Additionally, the fact that Pylint fails to reliably analyse the usage of prominent ML libraries, provides a major obstacle to the adoption of Continuous Integration (CI) in the development environment of ML systems. If a static code analysis tool produces too many false positives, it will be noisy and counterproductive \cite{vassallo2020asats}. Thus, other important true positives may be overlooked.

\highlight{Additionally, the fact that Pylint fails to reliably analyse whether prominent ML libraries are used correctly, provides a \textbf{major obstacle to the adoption of Continuous Integration (CI)} in the development environment of ML systems.}


\textbf{Warning} --
In this category, we found that one project (\codeinline{kaggle_rsna2019_3rd_solution}) was responsible for 13917 of all 26307 \codeinline{unused-wildcard-import} messages, with 53 \codeinline{wildcard-import} messages. Since \codeinline{unused-wildcard-import} are emitted per unused function imported with a wildcard import, this means that there were on average 263 unused imports per wildcard import in this project. Notably, most of these messages were also (contained in) instances of duplicate code. Such unused wildcard imports pollute a module's namespace with the names of all imported functions, meaning there is a greater chance of (accidentally) redefining an outer name. Additionally, wildcard imports may also have unintended side-effects that can be very difficult to debug.

The tendency towards using wildcard imports may stem from the prototypical and experimental nature of ML projects, combined with the fact that it is simply easier for the developer to import everything from a library and use whatever they need, rather than import functions individually. Dead experimental codepaths as found by \citet{sculley2015hidden} of which the imports still remain, can also be a cause of bad import management.


As for the many \codeinline{bad-indentation} messages, these were dominated primarily by DeepMind projects that were using a different convention for indentation width, namely two spaces instead of four. This is not surprising since indentation width is a preference, where the PEP8 style guide\footnote{\url{https://www.python.org/dev/peps/pep-0008/}} prescribes four spaces, but others such as Google's TensorFlow style guide\footnote{\url{https://www.tensorflow.org/community/contribute/code_style}} prescribe two spaces.

\textbf{Refactor} --
We found that \codeinline{duplicate-code} is the most commonly reported refactoring opportunity. Having manually inspected a random subset of these messages and where they occur, we have noticed that these are primarily caused by ML developers having multiple permutations of similar ML models to perform the same task. Each model (experimental codepath) then uses a slightly different underlying algorithm or slightly different parameters and are each defined in their own file, likely in an attempt to find the best performing one. Yet instead of identifying the commonalities between these different models and abstracting them into modules that can be reused across their codebase, ML developers seem to prefer simply copy-pasting the files. However, more research into code reuse and duplication in ML code is required to truly understand this phenomenon and how it can be prevented.

\highlight{\textbf{Code duplication is common in ML}, but calls for more extensive research to truly understand to what extent and for what reasons this occurs, and how it can be avoided.}

Regarding the high prevalence of the \codeinline{too-many-arguments} and \codeinline{too-many-locals} messages, it is congruent with previous work which shows that data science projects contain significantly more instances of these than traditional software projects~\cite{simmons2020dsstandards}. \citet{simmons2020dsstandards} also notes that these messages are related: function arguments namely also count as local function variables. By default, a \codeinline{too-many-arguments} message is emitted when a function or method takes more than five arguments, while \codeinline{too-many-locals} is emitted when a function or method contains more than 15 local variables. A possible cause for their prevalence, as \citet{simmons2020dsstandards} note, are \textit{``models with multiple hyperparameters that are either hard-coded as variables in the function definition or passed to the function as individual parameters rather than being stored in a configuration object.''}

\textbf{Convention} --
While line length violations stem primarily from developer preference, invalid and improper naming is a problem not exclusive to Python \cite{lacerda2020smellsSLR}. Pylint by default emits a \codeinline{invalid-name} message when it finds names that do not comply to PEP8, i.e. are improperly capitalised or less than three characters long (except in the case of inline variables). Indeed, shorter identifier names do take longer to comprehend than whole words \cite{hofmeister2019shortids}, but \citet{simmons2020dsstandards} aptly notes that this may not necessarily be the case in DS and ML code due to its heavily mathematical background. Thus, ML practitioners may find it easier to comprehend the details of a piece of ML code written so that it resembles the notation of the underlying mathematical model, including the names of the identifiers. Future research on this subject will have to show how this affects the readability of ML code.

\highlight{\textbf{The PEP8 convention} for identifier naming style \textbf{may not always be applicable in ML code} due to its resemblance with mathematical notation. Future research is required to investigate how this affects the readability of ML code.}





\subsection{Problems installing project dependencies} \label{sec:discussion:dependencies}

Setting up the projects' codebases as detailed in Section \ref{sec:methodology:setting-up} was not a trivial task. While 42 out of 74 projects did have a requirements file in their repository that installed without a hitch out of the box, there were 32 projects where a \codeinline{requirements.txt} had to be generated or had to manually be created from inspecting the repository or manually modified from what was already in the repository.
Furthermore, there were 13 projects that required installing extra dependencies after installing those in the requirements file. One of these projects had a valid requirements file -- and specified the extra dependencies in their ReadMe -- but the 12 others did not.

As for the projects for which a requirements file had to be manually created or modified, we made a few observations as to why this was needed. First, some projects did not contain a requirements file at all, but did specify instructions in the ReadMe, e.g., the \codeinline{yolact_edge} project.

Secondly, some projects had simply made a small mistake in their manual maintenance of the requirements file, as was the case with the \codeinline{navigan} project. The project authors fixed the mistake less than a day after we filed an issue on their GitHub\footnote{\url{https://github.com/yandex-research/navigan/issues/1}} about it.

Thirdly, some projects were relying on custom Docker containers for their runtime environment. These projects, e.g. \codeinline{kaggle-imaterialist}, maintain a Dockerfile in their repository (sometimes with an additional requirements file) in which the project's dependencies are installed, often without specifying exact dependency versions.

Finally, and most commonly (especially with the Kaggle projects), projects would contain a \codeinline{requirements.txt} file that was likely the result of running \codeinline{pip freeze} -- a shell command that lists all the packages installed in the current Python environment, including their respective dependencies, along with their exact versions. However, the are three problems with this approach:

\begin{itemize}[leftmargin=*]
    \item \textbf{Difficult to maintain} -- Since \codeinline{pip freeze} lists \textit{all} direct, indirect, runtime and development dependencies, without distinction, in alphabetical order, we conjecture that it is difficult for maintainers to assess whether a certain dependency can safely be upgraded without breaking their code or breaking any of their dependencies.
    \item \textbf{May result in unresolvable dependencies} -- The resulting requirements file may contain dependencies sourced from different dependency management tools and package indexes. These dependencies may have slightly different package names across package indexes or have only published certain versions to e.g. Conda's package index, but not to PyPI. There were also projects that depended on pre-release versions of certain libraries that are no longer available on PyPI (e.g., older nightly versions of Tensorflow packages).
    \item \textbf{May include unrelated dependencies} -- Especially if the user is not installing their dependencies into a virtual environment, then the resulting requirements file may also include unnecessary, unrelated (and potentially unresolvable) Python dependencies, such as those used by their operating system or those used in other projects. For example, the \codeinline{side_effects_penalties} in the DeepMind research repository depends on \codeinline{youtube-dl} (even though the project has nothing to do with videos), as well as some dependencies from the operating system level such as \codeinline{python-apt}, \codeinline{python-debian} and \codeinline{ufw}. The inclusion of the latter dependencies directly indicates that the project author was not using a virtual environment, but was instead using \codeinline{sudo pip install} to install all of their Python dependencies.
\end{itemize}

\highlight{We have found serious issues with the specification of dependencies that present a \textbf{major threat to the reproducibility and maintainability of Python ML projects}. Further research needs to be undertaken to help ML practitioners avoid issues in the dependency management of their projects.}



\section{Threats to Validity}

\subsection{Validity of the dataset}
Our dataset may not yet be fully representative of the real-world state of ML code, as it currently only contains open-source ML projects, Therefore, in future research, we want to collect a dataset of closed-source ML projects from the industry, such as ING's AI-driven FinTech industry. We will use this both to compare the prevalence of code smells in these closed-source industry projects with that of open-source projects as presented in this paper, as well as to make our dataset more representative of the real-world state of ML and AI projects. We will also explore adding projects from the dataset published by \citet{biswas2019dsdataset}.

Furthermore, we currently do not perform any analysis on the code quality of Jupyter Notebooks, even though they are very popular and have emerged as a de facto standard for data scientists \cite{perkel2018jupyter}. This was deliberate, as Pylint currently does not support directly analysing the Python code in Jupyter Notebook files and we wanted to avoid applying double standards to pure Python code and notebook Python code by extracting the notebook code into pure Python files. However, given their popularity, we do intend to perform future research on the code quality and linting of Jupyter Notebook code.


\subsection{Validity of Pylint}
Due to its dynamically typed nature, linting Python code is notoriously difficult \cite{simmons2020dsstandards,chen2018pysmells}. It is therefore no surprise that Pylint contains bugs and limitations that cause false positives and false negatives. Pylint's issue tracker on GitHub also reports 165 open and 501 closed issues regarding false positives\footnote{See \url{https://github.com/PyCQA/pylint/issues?q=is\%3Aissue+false+positive}} as of January 19th 2021. We have also noticed some of these shortcomings for ourselves during this research, as we have discussed in Section \ref{sec:discussion:code-smells}.
We mitigate this threat by manually checking a subset of projects to analyse potential false positives.

\section{Conclusion}

In this study we investigated the prevalence of code smells in ML projects. We gathered a dataset of 74 ML projects, ran the static analysis tool Pylint on them and collected the distribution of Pylint messages per category per project (\autoref{tab:results:msgs-per-category}), the top 10 code smells in these projects overall (\autoref{tab:results:top10-all}), and the top 20 code smells per category (\autoref{tab:results:top20-per-category}).

Moreover, by performing a manual analysis of a subset of the detected smells, we have found that code duplication is common in ML, but does require further research to understand to what extent this occurs and how it can be avoided. We also found that the PEP8 convention for identifier naming style may not always be applicable in ML code due to its resemblance with mathematical notation. This calls for additional research on how it affects the readability of ML code.

Most importantly, however, we have found serious issues with the specification of dependencies that present a major threat to the reproducibility and maintainability of Python ML projects. Furthermore, we found that Pylint produces a high rate of false positives on import statements and thus cannot reliably check for correct usage of imported dependencies, including prominent ML libraries such as PyTorch. Both of these problems also provide a major obstacle to the adoption of CI in ML systems. Further research needs to be undertaken to help ML practitioners avoid issues in the dependency management of their projects.




\renewcommand*{\bibfont}{\footnotesize}
\bibliographystyle{IEEEtranN}
\bibliography{refs}
\end{document}